\documentclass[fleqn,usenatbib,useAMS]{mnras}

\usepackage{newtxtext,newtxmath}

\usepackage[T1]{fontenc}

\DeclareRobustCommand{\VAN}[3]{#2}
\let\VANthebibliography\thebibliography
\def\thebibliography{\DeclareRobustCommand{\VAN}[3]{##3}\VANthebibliography}

\usepackage{graphicx}	
\usepackage{amsmath}	
\usepackage{orcidlink}

\newcommand{\msun}{{\rm M}_{\sun}}
\newcommand{\lsun}{{\rm L}_{\sun}}

\title[Debye cooling]{Debye cooling of ultramassive white dwarfs}

\author[D. Barba-González and S.Ginzburg]{
David Barba-González$^{\orcidlink{0000-0002-1518-5584}}$\thanks{E-mail: \href{mailto:david.barba-gonzalez@mail.huji.ac.il}{david.barba-gonzalez@mail.huji.ac.il}}
and Sivan Ginzburg$^{\orcidlink{0000-0002-3751-4553}}$
\\
Racah Institute of Physics, The Hebrew University, Jerusalem 9190401, Israel
}

\date{Accepted XXX. Received YYY; in original form ZZZ}

\pubyear{\the\year{}}

\begin{document}
\label{firstpage}
\pagerange{\pageref{firstpage}--\pageref{lastpage}}
\maketitle

\begin{abstract}
Recent observations of cold and massive white dwarfs are beginning to probe Debye cooling: an accelerated cooling phase driven by the vanishing heat capacity $C_V(T)$ of the crystallized interiors at low temperatures $T$, due to the quantum nature of phonon oscillations. We analyse this regime by calculating, both analytically and numerically, the instantaneous cooling time $\tau(L)\equiv -\textrm{d}t/\textrm{d}\ln L$, which is a proxy for the number density of white dwarfs at a given luminosity $L$. We find that cooling accelerates ($\tau$ shortens) at $L\lesssim 10^{-4}\,\lsun$ due to a combination of Debye cooling and the simultaneous convective coupling of the degenerate core to the surface, with neither process causing significant acceleration on its own. Specifically, strong magnetic fields can inhibit convective coupling, leading to an almost flat $\tau\propto L^{1/7}$ instead. We parametrize the uncertainty in the solid microphysics by generalizing the Helmholtz free energy such that $C_V\propto T^\delta$ at low temperatures. The currently observed decreasing luminosity function in the Debye regime is tentatively consistent with $\tau(L)$ of non-magnetic white dwarfs with $\delta=3$, corresponding to the standard Debye model. Larger samples could probe, in addition to magnetism, deviations $\delta<3$, which diverge appreciably before the Galactic disc's age $t\approx 10\textrm{ Gyr}$ for masses $M\gtrsim 1.0\,\msun$. For $\delta>3$, in contrast, the cooling curves converge because the heat capacity becomes dominated by the degenerate electrons $C_V\propto T$.
\end{abstract}




\section{Introduction}
\label{sec:intro}

White dwarfs have long been recognized as useful cosmic clocks \citep{Schmidt1959}. The heat stored in their degenerate cores is radiated away over billions of years by photon diffusion through their ideal gas envelopes at a predictable rate \citep{Mestel1952}, relating a white dwarf's luminosity and effective temperature to its age. Using this relation, the observed white dwarf luminosity distribution can determine the age of stellar clusters, our Galaxy, and even the Universe \citep{Winget1987,HansenLiebert2003,WingetKepler2008}.

However, the high densities of white dwarf cores complicate their cooling once they reach sufficiently low temperatures \citep[see][for reviews]{Fontaine2001,Althaus2010,Renedo2010}. Specifically, the cores crystallize into an ionic lattice \citep{VanHorn1968}, with the phase transition and associated processes leading to a cooling delay \citep[e.g.][]{Tremblay2019,Bauer2023}, which in some cases may amount to billions of years \citep[a result of neon gravitational energy release in particular; see][]{Isern1991,Segretain1994,Cheng2019,Blouin2021,Camisassa2021,Bedard2024}. 
At even lower temperatures, the heat capacity of the crystal lattice vanishes due to the quantum nature of its phonon oscillations: below the Debye temperature, some oscillation modes can no longer be thermally excited, losing their capacity to store heat \citep{MestelRuderman1967}. This leads to an accelerated cooling of the white dwarf \citep{LambVanHorn1975,Althaus2007}. 

Typical white dwarfs with masses $M\approx 0.6\,\msun$ \citep[e.g.][]{OBrien2024} reach the Debye regime only at cooling times longer than the age of the Galaxy \citep{ShavivKovetz1976,Althaus2010}. Ultramassive white dwarfs with $M\gtrsim 1.0\,\msun$, on the other hand, are much rarer, but they cool faster and reach the Debye regime earlier \citep{Althaus2007,Camisassa2019}. Such ultramassive white dwarfs probably formed through two channels with similar contributions \citep{Cheng2020,Fleury2022,Kilic2023,Jewett2024}: single stellar evolution of progenitors in a limited initial mass range \citep{Nomoto1984,Siess2010,Doherty2015,Kilic2023single,Miller2023,Limongi2024}, and mergers of two lower mass white dwarfs \citep{Segretain1997,Shen2012,Hollands2020,Caiazzo2021,Schwab2021,Kawka2023,Kilic2024}. The core composition of ultramassive white dwarfs also seems to vary between carbon--oxygen (CO) and oxygen--neon (ONe), depending on whether carbon is ignited upon formation \citep{TruranLivio1986,Dominguez1996,Bauer2020,Althaus2021,Blatman2025,Arseneau2026}.

Until recently, cold and massive white dwarfs, which are expected to be in the Debye regime, have remained elusive, with the lack of observations attributed to the rapid cooling in this regime down to undetectable luminosities \citep{Kilic2025}. Notable exceptions are a single magnetic DA (i.e. hydrogen atmosphere) ultramassive white dwarf in the Debye regime \citep{Kilic2026}, and a population of low effective temperature infrared-faint (IR-faint) white dwarfs,\footnote{See \citet{Saumon2022} for a brief explanation of this apparent paradox in the spectrum.} a subset of which are sufficiently massive to be in the Debye regime \citep{Bergeron2022}. These new detections open the first observational window for probing the physics of Debye cooling in white dwarf stars. 

A white dwarf's cooling rate is determined not only by the heat capacity of its core, but also by how this heat is transported to the surface. As white dwarfs cool down, they develop surface convection zones that gradually expand inward \citep{Bohm1968,vanHorn1970}. Initially, this outer convective region does not affect the white dwarf's cooling rate, because the interior degenerate core is insulated from the envelope by an intermediate radiative layer. Only once the convective zone directly couples to the degenerate core -- eliminating the radiative bottleneck -- can heat be transported more efficiently by convection \citep{Fontaine2001}. 

In ultramassive white dwarfs, convective coupling and Debye cooling are almost simultaneous \citep{LambVanHorn1975,Fontaine2001,Bergeron2022}, and they entangle non-trivially to determine the cooling rate \citep{DAntonaMazzitelli1989,Obertas2018}. To complicate the picture further, strong magnetic fields $B\gtrsim 1\textrm{ MG}$ can inhibit convective energy transfer and decouple the core from the envelope \citep{Tremblay2015,Ginzburg2024}. Recent volume-limited surveys found that such strong fields are common in cold white dwarfs \citep{BagnuloLandstreet2022,Moss2025}, with double white dwarf mergers \citep{GarciaBerro2012}, crystallization dynamos \citep{Isern2017,Camisassa2022,BlatmanGinzburg2024b}, and the diffusion of magnetic fields from a previous evolutionary stage \citep{Camisassa2024,Castro-Tapia2026} all invoked as possible origins for ultramassive magnetic white dwarfs in particular (which are already magnetic at an early age). 

Here, motivated by the first observations of white dwarfs in the Debye regime and by the recent discovery that strong magnetic fields are common in old and in massive white dwarfs, we revisit the theory of Debye cooling in ultramassive white dwarfs. Specifically, we disentangle the effects of Debye cooling from convective coupling (which might be averted in magnetic white dwarfs), and also examine to what extent samples like \citet{Bergeron2022} can probe the microphysics of the phonon heat capacity in crystallized white dwarfs. 

The remainder of this paper is organized as follows. In Section \ref{sec:capacity}, we present our microscopic model for the phonon heat capacity. 
In Section \ref{sec:cooling}, we estimate white dwarf cooling rates in different regimes of the heat capacity and transport using a simple analytical model.   
We discuss our numerical results and their relation to the observations in Section \ref{sec:numerical}, and summarize our conclusions in Section \ref{sec:conclusions}.

\section{Crystal heat capacity}
\label{sec:capacity}

Once crystallized, the white dwarf's core can be modelled as a lattice of coupled quantum harmonic oscillators, with collective vibration modes -- each with a frequency $\omega$ and energy states $E_j=\hbar\omega(j+1/2)$ with an integer $j$ ($\hbar$ is the reduced Planck constant). The canonical partition function of each vibration mode is given by summing over all the energy states
\begin{equation}\label{eq:partition}
    Z_\omega=\sum_{j=0}^{\infty}{e^{-\beta E_j}}=\sum_{j=0}^{\infty}{e^{-\beta \hbar\omega j}}=\left(1-e^{-\beta\hbar\omega}\right)^{-1},
\end{equation}
where $\beta\equiv (kT)^{-1}$, $k$ is Boltzmann's constant, $T$ is the temperature, and where the thermal energy is measured relative to the ground state $\hbar\omega/2$. The total partition function $Z$ is given by multiplying the individual ones $Z_\omega$, such that
\begin{equation}\label{eq:lnZ}
\ln Z=\int\limits_0^{\omega_{\rm D}}g(\omega){\rm d}\omega\ln Z_\omega=-\int\limits_0^{\omega_{\rm D}}g(\omega){\rm d}\omega\ln\left(1-e^{-\beta\hbar\omega}\right),
\end{equation}
where $g(\omega)$ is the density of states (vibration modes) and $\omega_{\rm D}$ is the maximal cut-off frequency introduced by Debye. The system's thermal energy is given by
\begin{equation}\label{eq:U}
    U=-\frac{\partial\ln Z}{\partial\beta}=\int_0^{\omega_{\rm D}}{\frac{\hbar\omega}{e^{\beta\hbar\omega}-1}g(\omega)}{\rm d}\omega.
\end{equation}
Equation \eqref{eq:U} can be understood as a sum of all phonons (the quantized excitations of the vibration modes) -- each with an energy $\hbar\omega$ -- with the average occupancy of each state given by the Bose--Einstein distribution (with a zero chemical potential, i.e. the Planck distribution). At high temperatures $kT\gg\hbar\omega$, each oscillation mode is excited to $E\approx(\hbar\omega)/(1+\beta\hbar\omega-1)=kT$. At low temperatures $kT\ll\hbar\omega$, on the other hand, the occupation number is exponentially small such that the mode is practically unable to store heat.  

For a single phonon branch (analogous to polarization) we choose a density of states of the form
\begin{equation}\label{eq:dos}
    g(\omega)=\frac{N\delta}{\omega_{\rm D}^\delta}\omega^{\delta-1}\propto\omega^{\delta-1},
\end{equation}
where $N$ is the total number of ions. In the standard Debye model, a linear dispersion relation yields $\delta=3$ in three dimensions (the wave vector defines a sphere with a radius proportional to $\omega$). We introduce here the generalization to other values of $\delta$ as a simple parametrization of the uncertainty in the phonon spectrum.
The normalization in equation \eqref{eq:dos} is chosen such that the total number of states (i.e. degrees of freedom) is equal to the number of ions in the lattice
\begin{equation}
    \int_0^{\omega_{\rm D}}g(\omega){\rm d}\omega=N.
\end{equation}
Substituting the density of states $g(\omega)$ from equation \eqref{eq:dos} into equation \eqref{eq:U} yields the thermal energy
\begin{equation}\label{eq:thermal_energy}
    U=NkTD_\delta(\theta),
\end{equation}
where $T_{\rm D}\equiv\hbar\omega_{\rm D}/k$ is the Debye temperature, $\theta\equiv T_{\rm D}/T$, and
\begin{equation}
    D_\delta(\theta)\equiv\frac{\delta}{\theta^\delta}\int_0^\theta{\frac{x^\delta}{e^x-1}}{\rm d}x
\end{equation}
is the Debye function of order $\delta$, which obviously converges only for $\delta>0$. 

\begin{figure}
	\includegraphics[width=\columnwidth]{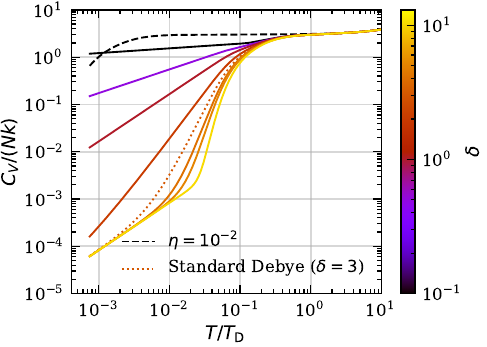}
    \caption{Our modified heat capacity for harmonic phonons, according to equation \eqref{eq:F_chabrier}, and degenerate electrons \citep[][where we assume pure carbon at a density $\rho=10^7\textrm{ g cm}^{-3}$ for illustration]{PotekhinChabrier2000}, as a function of $T/T_{\rm D}=\theta^{-1}$, where $T_{\rm D}$ is the Debye temperature. The nominal Debye heat capacity ($\delta=3$, $\eta=1$) is given by the dotted line, with the solid lines (colour coded according to $\delta$) dropping at different rates $C_V\propto T^\delta$ in the Debye regime. For high values of $\delta$, the heat capacity becomes dominated by the electronic $C_V\propto T$ at low temperatures. The Dulong--Petit limit is reached by taking $\delta=3$ and $\eta\ll 1$, effectively shifting the Debye regime to arbitrarily low temperatures (dashed black line).}
    \label{fig:cv}
\end{figure}

At high temperatures $T\gg T_{\rm D}$, $D_\delta(\theta\ll 1)\approx 1$ such that $U\approx NkT$. In other words, each vibration mode is excited to $kT$ and the heat capacity is $C_V=\partial U/\partial T=Nk$, reproducing the Dulong--Petit law (according to which $C_V=3Nk$ due to the three polarizations). At low temperatures $T\ll T_{\rm D}$, $D_\delta(\theta\to\infty)\sim\theta^{-\delta}$ such that $C_V\sim Nk (T/T_{\rm D})^\delta$. In three dimensions $\delta=3$, the heat capacity is reduced by $(T/T_{\rm D})^3$ because only an inner ball of radius $\omega=kT/\hbar$ (out of a total radius $\omega_{\rm D}$) of the oscillations can be excited. We use different values of $\delta>0$ to model a shallower or steeper decline of $C_V$ for $T<T_{\rm D}$ (Fig. \ref{fig:cv}).

Numerical equations of state are typically formulated in terms of the Helmholtz free energy $F=-kT\ln Z$, from which all other thermodynamic quantities can be derived consistently \citep[e.g.][]{Jermyn2021}. By substituting $g(\omega)$ from equation \eqref{eq:dos} into equation \eqref{eq:lnZ}, and after integration by parts:  
\begin{equation}
    \frac{F}{NkT}=\ln\left(1-e^{-\theta}\right)-\frac{1}{\delta}D_\delta(\theta).
\end{equation}
More precisely, we follow \citet{Chabrier1992} and choose

\begin{equation}\label{eq:F_chabrier}
    \frac{F}{NkT}=2\ln\left(1-e^{-\eta\alpha\theta}\right)-\frac{2}{\delta}D_\delta(\eta\alpha\theta)+\ln\left(1-e^{-\eta\gamma\theta}\right),
\end{equation}
where the factor of 2 represents two transverse branches, and the last term represents an additional longitudinal branch that is assumed to follow an Einstein $g(\omega)$ instead of a Debye one (i.e. an approximately flat dispersion relation instead of a linear one).\footnote{Our model inherently assumes three phonon branches, such that setting $\delta\neq 3$ does not fully change the dimensionality. As a result, the Dulong--Petit limit is still $C_V\to 3Nk$ for $T\gg T_{\rm D}$.} In an Einstein solid, all vibrations are at the same frequency $\omega=\omega_{\rm D}$, such that the integration in equation \eqref{eq:lnZ} is trivial. We do not modify this term (except for the $\eta$ parameter which is explained below) because the Einstein heat capacity vanishes exponentially at $T\ll T_{\rm D}$ in any case -- much faster than the Debye $C_V\propto T^\delta$.

The final component in our model is the cut-off frequency $\omega_{\rm D}$. In the standard Debye model this frequency is similar to the ion plasma frequency, which also describes the oscillation of individual ions in a lattice: 
\begin{equation}\label{eq:frequency}
    \omega_{\rm D}=\left(\frac{4\upi n q^2}{m}\right)^{1/2},
\end{equation}
where $m$, $q$, and $n$ are the ion mass, electric charge, and number density, respectively. Since individual ion oscillations represent the shortest wavelength for collective vibrations, they set the maximal phonon frequency. We follow \citet{Chabrier1993} and add the factors $\alpha\approx 0.4$ and $\gamma\approx 0.9$ to equation \eqref{eq:F_chabrier}, for a better agreement with the numerically computed phonon spectrum. We introduce an additional adjustable parameter $\eta$ to equation \eqref{eq:F_chabrier} that controls the effective Debye frequency and temperature for all the branches.\footnote{\citet{Chabrier1992} and \citet{Chabrier1993} denote $\eta\equiv T_{\rm D}/T$, whereas we follow the notation of \citet{Baiko2001} $\theta\equiv T_{\rm D}/T$, and introduce $\eta$ as an additional prefactor.} 

In summary, our expression for the free energy has two free parameters $\eta$ and $\delta$, which control the effective Debye temperature and the rate at which the heat capacity vanishes at lower temperatures, respectively. For $\eta=1$ and $\delta=3$ our model is identical to \citet{Chabrier1993}. We note that the \textsc{mesa} stellar evolution code \citep{Paxton2011,Paxton2013,Paxton2015,Paxton2018,Paxton2019,Jermyn2023}, which we use to compute white dwarf cooling, employs the Skye equation of state \citep{Jermyn2021} and specifically the \citet{Baiko2001} free energy to describe the thermodynamics of harmonic oscillations in the lattice. \citet{Baiko2001} heuristically generalized \citet{Chabrier1992} by replacing the Debye function with a rational function (i.e. the ratio of two polynomials), which can be fit more accurately to numerically computed phonon spectra. In this paper, we implement instead our different generalization of \citet{Chabrier1992}, i.e. equation \eqref{eq:F_chabrier}, because it is physically more tractable. For the default parameters, the two expressions agree to within 10 per cent \citep{Baiko2001}, such that the conclusions of our paper are unaffected.

As demonstrated in Fig. \ref{fig:cv}, our generalized expression for the solid (crystal) ion free energy enables us to experiment with a shallower power law decay of the heat capacity in the Debye regime by adjusting $0<\delta<3$, up to the Dulong--Petit limit (i.e. no decay), which is reached in practice by setting $\eta\to 0$ (which controls the Einstein branch as well). Experimenting with steeper power laws $\delta>3$, on the other hand, is limited, because the heat capacity quickly becomes dominated by the degenerate electrons $C_V\propto T$ \citep{PotekhinChabrier2000,Fontaine2001,Obertas2018}.\footnote{Intuitively, only a small fraction $\sim T/T_{\rm F}$ of the electrons -- those at the top of the Fermi sea -- are thermally excited, where $T_{\rm F}$ is the Fermi temperature.}
We note that there are other contributions to the solid free energy, in addition to degenerate electrons and harmonic phonons, such as anharmonic corrections, electron--ion interactions, and mixing between the different ion species \citep[see][for the full expression used in \textsc{mesa}]{Jermyn2021}. In this paper, we modify only the harmonic phonon term in \textsc{mesa} according to equation \eqref{eq:F_chabrier}, with all other contributions preserved. 

\section{Cooling rate}
\label{sec:cooling}

In the classical cooling model of \citet{Mestel1952}, the white dwarf is composed of a degenerate and isothermal (thanks to the heat conductivity of degenerate electrons) core at a temperature $T_{\rm c}$, surrounded by a thin radiative ideal gas envelope that regulates its cooling. The transition between these two layers is given by equating the electron degeneracy pressure $P\propto\rho^{5/3}$ to the ideal gas pressure $P\propto \rho T_{\rm c}$, such that electrons become degenerate at a mass density
\begin{equation}\label{eq:rho_deg}
    \rho_{\rm deg}\propto T_{\rm c}^{3/2}.
\end{equation}
We note that even in ultramassive white dwarfs, where the degenerate electrons at the centre are relativistic, i.e. $P\propto \rho^{4/3}$, the transition to ideal gas is at lower densities, where the electrons are non-relativistic \citep{BlatmanGinzburg2024b}. By assuming Kramers' opacity law for the gas directly above the core -- which is the bottleneck for cooling -- \citet{Mestel1952} derived the white dwarf's luminosity
\begin{equation}\label{eq:mestel}
    L\propto M T_{\rm c}^{7/2},
\end{equation}
irrespective of its radius $R$; see \citet{VanHorn1971} and \citet{BlatmanGinzburg2024b} for details of the derivation.

The white dwarf begins to crystallize when the plasma coupling parameter at its centre $\Gamma\propto \rho_{\rm c}^{1/3}T_{\rm c}^{-1}$ exceeds a critical value, with $\rho_{\rm c}\sim MR^{-3}$ denoting the central density \citep{VanHorn1968,Jermyn2021}. Using equation \eqref{eq:mestel}, this corresponds to a luminosity
\begin{equation}\label{eq:L_cryst}
    L_{\rm cryst}\propto M\rho_{\rm c}^{7/6}\propto M^{13/6}R^{-7/2}\propto M^{10/3}\left[1-\left(\frac{M}{M_{\rm ch}}\right)^{4/3}\right]^{-7/4},
\end{equation}
where we use the approximate mass--radius relation \citep{TruranLivio1986,Hansen2004}
\begin{equation}\label{eq:mass_rad}
   R\propto M^{-1/3}\left[1-\left(\frac{M}{M_{\rm ch}}\right)^{4/3}\right]^{1/2}, 
\end{equation}
with $M_{\rm ch}\approx 1.46\,\msun$ denoting the Chandrasekhar mass. Similarly, the white dwarf reaches the Debye regime when $T_{\rm c}$ drops below the Debye temperature at the centre, which is given by equation \eqref{eq:frequency}: $kT_{\rm c}=\hbar\omega_{\rm D}\propto \rho_{\rm c}^{1/2}$. Using equations \eqref{eq:mestel} and \eqref{eq:mass_rad}, the corresponding luminosity is
\begin{equation}\label{eq:L_Debye}
    L_{\rm Debye}\propto M\rho_{\rm c}^{7/4}\propto M^{9/2}\left[1-\left(\frac{M}{M_{\rm ch}}\right)^{4/3}\right]^{-21/8}.
\end{equation}

However, the \citet{Mestel1952} model, specifically equation \eqref{eq:mestel}, breaks down before the Debye regime is reached because of convective coupling. \citet{Yaakovyan2025} demonstrated that, after an initial build-up phase, the base of the convection zone settles at an almost constant temperature. Using their equations (4) and (5), the density at the base of the convection zone increases as the white dwarf cools as
\begin{equation}\label{eq:rho_conv}
    \rho_{\rm conv}\propto\left(\frac{M}{L}\right)^{1/2},
\end{equation}
where the radius cancels out.
The density at the top of the degenerate core, on the other hand, decreases as
\begin{equation}\label{eq:rho_deg_L}
    \rho_{\rm deg}\propto \left(\frac{L}{M}\right)^{3/7},
\end{equation}
using equations \eqref{eq:rho_deg} and \eqref{eq:mestel}. The inwardly expanding convection zone and the outwardly expanding degenerate core are bound to couple ($\rho_{\rm conv}=\rho_{\rm deg}$) at a luminosity
\begin{equation}\label{eq:L_coup}
    L_{\rm coup}\propto M.
\end{equation}
As we show in Section \ref{sec:numerical} \citep[see also][]{Fontaine2001}, in ultramassive white dwarfs the order of events is $L_{\rm cryst}>L_{\rm coup}>L_{\rm Debye}$, such that the \citet{Mestel1952} model and equation \eqref{eq:mestel} in particular must be amended in order to study the Debye cooling regime.\footnote{Crystallization and Debye cooling commence when white dwarfs cool down to $T_{\rm c}\propto\rho_{\rm c}^{1/3}$ and $T_{\rm c}\propto\rho_{\rm c}^{1/2}$, respectively. In principle, the densest white dwarfs with $M\to M_{\rm ch}$ may therefore reach the quantum cooling regime before crystallizing \citep{Chabrier1993}. In practice, crystallization precedes Debye cooling for the $M\lesssim 1.2\,\msun$ white dwarfs that we simulate in Section \ref{sec:numerical}.} 

In order to model white dwarf cooling after convective coupling, we generalize equation \eqref{eq:mestel} similarly to \citet{DAntonaMazzitelli1989} and \citet{Obertas2018}:
\begin{equation}\label{eq:epsilon}
L\propto T_{\rm c}^\epsilon,   
\end{equation}
where we disregard the dependence on $M$ for the rest of this section (i.e. we focus on the evolution of a white dwarf with a given mass). Whereas $\epsilon\approx 7/2$ before convective coupling \citep{Mestel1952}, it becomes considerably smaller once the convection zone reaches the degenerate core -- eliminating the radiative bottleneck \citep{DAntonaMazzitelli1989,Tassoul1990,Fontaine2001}. Intuitively, convection flattens the temperature gradient between the core $T_{\rm c}$ and the photosphere $T_{\rm eff}\propto L^{1/4}$, increasing the luminosity at a fixed $T_{\rm c}$. 

Using equations \eqref{eq:thermal_energy} and \eqref{eq:epsilon}, we can finally estimate the time $\tau$ that the white dwarf spends at a given luminosity \citep[see also][]{DAntonaMazzitelli1989}
\begin{equation}\label{eq:tau}
    \tau\sim\frac{U}{L}\propto T_{\rm c}^{\delta+1-\epsilon}\propto L^{(\delta+1-\epsilon)/\epsilon},
\end{equation}
where in this section we consider only the power-law limit of the heat capacity $C_V\propto T_{\rm c}^\delta$ (see Section \ref{sec:capacity}). 
More precisely,
\begin{equation}\label{eq:tau_def}
    \tau(L)\equiv -\frac{{\rm d}t}{{\rm d}\ln L}
\end{equation}
is defined as the time $t$ spent in a logarithmic luminosity bin and it is proportional -- up to the initial mass function and the star formation rate -- to the number of white dwarfs with a given luminosity \citep[the luminosity function, see][]{Fontaine2001,GarciaBerroOswalt2016}.\footnote{If the exponent in equation \eqref{eq:tau} is negative, it is easy to see that $\tau(L)$ is similar to the total cooling time $t$ required to reach a luminosity $L$.}

For a radiative envelope, $\epsilon=7/2$, such that equation \eqref{eq:tau} predicts $\tau\propto L^{-5/7}$ in the Dulong--Petit limit \citep[$\delta=0$, corresponding to][]{Mestel1952}, and an almost flat $\tau\propto L^{1/7}$ in the Debye regime ($\delta=3)$. However, equation \eqref{eq:tau} indicates that the heat capacity (parametrized by $\delta$) and transport (i.e. convective or radiative, parametrized by $\epsilon$) are tangled together. Specifically, convective coupling and Debye cooling, which are almost simultaneous in ultramassive white dwarfs \citep{Fontaine2001}, exacerbate each other ($\delta$ increases while $\epsilon$ decreases). Quantitatively, the difference in the exponent of $\tau(L)$ between different heat capacities $\delta_{1,2}$ (e.g. Debye versus Dulong--Petit)
\begin{equation}
    \left.\frac{{\rm d}\ln\tau}{{\rm d}\ln L}\right|_{\delta_2}-\left.\frac{{\rm d}\ln\tau}{{\rm d}\ln L}\right|_{\delta_1}=\frac{\delta_2-\delta_1}{\epsilon}
\end{equation}
is larger for convectively coupled envelopes, which have a smaller $\epsilon<7/2$.
Similarly, the difference between radiative and convectively coupled envelopes (represented by different $\epsilon_{1,2}$)
\begin{equation}
    \left.\frac{{\rm d}\ln\tau}{{\rm d}\ln L}\right|_{\epsilon_2}-\left.\frac{{\rm d}\ln\tau}{{\rm d}\ln L}\right|_{\epsilon_1}=(\delta+1)\left(\frac{1}{\epsilon_2}-\frac{1}{\epsilon_1}\right)
\end{equation}
is much larger in the Debye regime ($\delta=3$) than in the Dulong--Petit regime ($\delta=0)$.

In summary, this simple analytical model demonstrates that Debye cooling and convective coupling are a combined phenomenon in ultramassive white dwarfs. However, strong magnetic fields $B\gtrsim 1\textrm{ MG}$, which are common in young ultramassive white dwarfs and in old lower-mass white dwarfs \citep{BagnuloLandstreet2022,Moss2025}, may open a radiative window above the core that decouples it from the convective envelope \citep{Ginzburg2024}. Such magnetic ultramassive white dwarfs enter the Debye regime and lose their heat capacity while still transporting the heat radiatively -- disentangling the two effects. In Section \ref{sec:numerical} we compute $\tau(L)$ more accurately using \textsc{mesa} for both magnetic and non-magnetic white dwarfs, and also test its sensitivity to uncertainties in the phononic heat capacity.   

\section{Numerical results}
\label{sec:numerical}

\subsection{White dwarf models}\label{sec:models}

We compute massive white dwarf models using the \textsc{mesa} (version r24.08.1) test suite \texttt{make\_co\_wd}, which evolves pre-main sequence stars with different initial masses through all stages of stellar evolution, yielding CO white dwarfs with $M\lesssim 1.0\,\msun$.\footnote{The \texttt{make\_co\_wd} and \texttt{make\_o\_ne\_wd} test suites simplify some stages of stellar evolution for numerical convenience.} As we are interested in ultramassive white dwarfs, we employ the procedure \texttt{relax\_mass\_scale}, which incrementally rescales the total mass of the star while keeping the chemical composition constant as a function of the fractional mass coordinate. We apply this procedure to the $0.8\,\msun$ model in order to produce our nominal grid of CO white dwarfs with masses $0.8\,\msun\lesssim M \lesssim 1.16 \,\msun$. We have verified that the rescaling procedure does not affect our results in the overlapping mass range $0.8 \,\msun \lesssim M\lesssim 1.0\,\msun$. Similarly, we use the \texttt{make\_o\_ne\_wd} test suite to form an initially $1.1\,\msun$ ONe white dwarf, which we then rescale using \texttt{relax\_mass\_scale} in order to compare CO and ONe white dwarfs with the same mass.\footnote{We use the larger \texttt{sagb\_NeNa\_MgAl.net} nuclear reaction network, instead of \textsc{mesa}'s standard \texttt{co\_burn\_plus.net}, improving the accuracy of the oxygen to neon ratio \citep{deGeronimo2022}.}

Our nominal white dwarfs have hydrogen-dominated outer atmospheres (we use the \texttt{replace\_element} procedure to create a hydrogen layer on top of the initially pure helium atmospheres of our ONe models), representing the majority of the ultramassive white dwarf population before it enters the Debye regime \citep{Kilic2025}. We note, however, that the \citet{Bergeron2022} IR-faint sample is best explained by mixed outer atmospheres: dominated by helium, with an uncertain fraction of hydrogen. We explore helium-dominated and mixed atmospheres in Appendix \ref{sec:appendixmixed} using the \texttt{replace\_element} procedure. However, \textsc{mesa}'s tabulated boundary conditions for such atmospheres reach down only to $T_{\rm eff}\approx 5000\textrm{ K}$ (compared to $T_{\rm eff}\approx 2000\textrm{ K}$ for hydrogen atmospheres), limiting our ability to probe the Debye regime \citep[see also][]{Bauer2019}.

\begin{figure}
    \centering
    \includegraphics[width=\columnwidth]{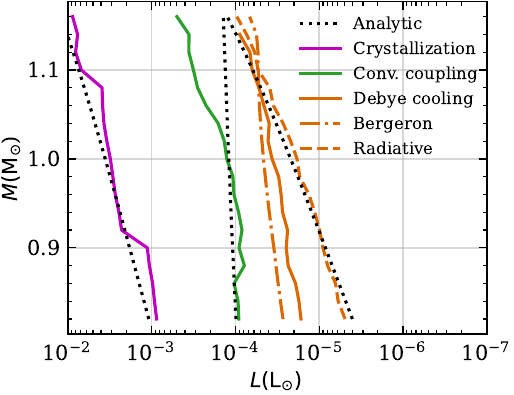}
    \caption{From left to right, solid lines show the onset of crystallization, convective coupling, and Debye cooling in the mass--luminosity plane for CO white dwarfs with hydrogen atmospheres. Dotted lines show the corresponding analytical predictions for $L_{\rm cryst}(M)$, $L_{\rm coup}(M)$, and $L_{\rm Debye}(M)$ from Section \ref{sec:cooling}. Specifically, equation \eqref{eq:L_Debye} fits the onset of Debye cooling only if convection is artificially turned off (dashed line), in compliance with the assumptions of \citet{Mestel1952}. Finally, the dot--dashed line corresponds to the onset of Debye cooling as shown in \citet{Bergeron2022}, based on the cooling sequences of \citet{bedard2020}.  
    }
    \label{fig:stagescooling}
\end{figure}

We cool our models (with the standard Debye heat capacity $\delta=3$, $\eta=1$) using the \texttt{wd\_cool\_0.6M} test suite, and present the relevant stages in Fig. \ref{fig:stagescooling}. For the entire mass range, $L_{\rm cryst}>L_{\rm coup}>L_{\rm Debye}$, with the onset of crystallization \citep[$\Gamma>170$ approximately, see][]{PotekhinChabrier2000,Jermyn2021} and Debye cooling \citep[$T_{\rm D}/T>20$, following][]{Althaus2007} at the centre reproduced by our analytical equations \eqref{eq:L_cryst} and \eqref{eq:L_Debye}, respectively. Convective coupling \citep[the electron chemical potential $>0$ at the base of the convection zone, see][]{Fontaine2001}, on the other hand, deviates from our analytical equation \eqref{eq:L_coup}, due to the approximation of a constant temperature at the base of the convection zone \citep[see also fig. 5 of][]{Yaakovyan2025}. More importantly, our analytical equation \eqref{eq:L_Debye} reproduces the onset of Debye cooling only when we artificially turn off convection in \textsc{mesa}. In contrast, both our nominal models (i.e. with convection included) and the models used by \citet{Bergeron2022} (except for the highest masses) enter Debye cooling earlier, illustrating the interplay between convective coupling and Debye cooling, which we discussed in Section \ref{sec:cooling}. We investigate this interplay further in Section \ref{sec:numerical_cooling}.

\subsection{Convective coupling and Debye cooling}\label{sec:numerical_cooling}

\begin{figure}
    \centering
    \includegraphics[width=\columnwidth]{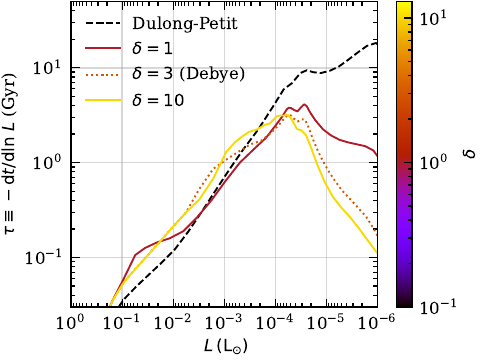}
    \caption{The instantaneous cooling time $\tau$ for a $1.0\,\msun $ CO white dwarf with a hydrogen atmosphere, assuming different models for the phonon heat capacity $C_V$ (see Fig. \ref{fig:cv}). These include the Dulong--Petit limit (dashed line), the standard Debye model ($\delta=3$, $\eta=1$; dotted line), and heat capacities that vanish at different rates in the Debye regime $C_V\propto T^\delta$ (solid lines).} 
    \label{fig:debye}
\end{figure}

We analyse Debye cooling using the instantaneous cooling time $\tau(L)$, which is defined in equation \eqref{eq:tau_def}. In Fig. \ref{fig:debye} we plot $\tau$ for our nominal $1.0\,\msun$ model (CO core, hydrogen atmosphere, convection included), while varying the phonon heat capacity $C_V$.\footnote{We edited the \texttt{skye\_coulomb\_solid.f90} source file inside the \texttt{eos/private/} directory of \textsc{mesa}.} At high luminosities $L\gtrsim 10^{-4}\,\lsun$, the cooling rate slows down (i.e. $\tau$ increases) over time, in agreement with \citet{Mestel1952}. This trend reverses as the star enters the Debye regime, and cooling accelerates \citep[i.e. $\tau$ decreases, see also][]{Fontaine2001}. The slope of $\tau(L)$ depends on the rate at which the phonon heat capacity vanishes $C_V\propto T^\delta$ for $\delta\lesssim 3$, whereas for higher values of $\delta$ the curves are almost indistinguishable from the standard Debye case ($\delta=3$), because the heat capacity becomes dominated by the degenerate electrons (see Section \ref{sec:capacity}). 

\begin{figure}
    \centering
    \includegraphics[width=\columnwidth]{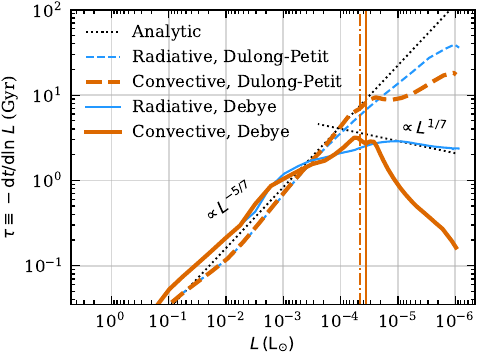}
    \newline
    \includegraphics[width=\columnwidth]{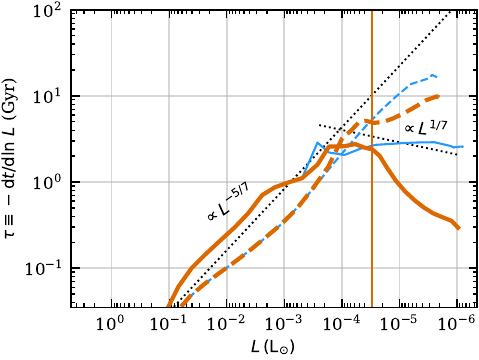}
    \caption{The instantaneous cooling time $\tau$ for $1.0\,\msun$ CO (top panel) and ONe (bottom panel, with qualitatively and quantitatively similar results) white dwarfs with hydrogen atmospheres. Our nominal model includes both convection and standard Debye cooling (thick solid line), whereas in the other models we artificially turn off either one or both of these effects. Dotted lines show the analytical limits from equation \eqref{eq:tau} for radiative envelopes ($\epsilon=7/2)$. Vertical lines mark the onset of Debye cooling $T_{\rm D}/T=20$ at the centre for our nominal model (solid) and in the \citet{Bergeron2022} CO model (dot--dashed), which roughly coincide with the peak in $\tau(L)$.}
    \label{fig:convection}
\end{figure}

However, as explained analytically in Section \ref{sec:cooling}, the $\tau(L)$ slope depends not only on the heat capacity, but also on the heat transport (i.e. convection or radiation, parametrized by $\epsilon$), since convective coupling slightly precedes the Debye regime in ultramassive white dwarfs (Fig. \ref{fig:stagescooling}). We demonstrate this interplay and disentangle Debye cooling from convective coupling in Fig. \ref{fig:convection}, by artificially turning off convection in \textsc{mesa} (for the entire white dwarf throughout its cooling). These fully radiative models ($\epsilon\approx 7/2$) agree with our analytical equation \eqref{eq:tau} in both the Dulong--Petit ($\delta=0$) and Debye ($\delta=3$) limits. Fig. \ref{fig:convection} demonstrates that the accelerated cooling (i.e. sharp drop in $\tau$) at $L\lesssim 10^{-4}\,\lsun$ is a combination of convective coupling and Debye cooling, and cannot be attributed to any one of these two effects individually.  

Specifically, previous studies have demonstrated that magnetic inhibition of convective coupling has a subtle effect on the cooling of typical-mass white dwarfs, which never reach the Debye regime \citep{Tremblay2015,Ginzburg2024}. Such a subtle effect is also evident in Fig. \ref{fig:convection}, when Debye cooling is averted by artificially forcing a Dulong--Petit heat capacity ($\eta=10^{-2}$; dashed lines). However, Fig. \ref{fig:convection} shows that convective coupling is a much more significant event in the cooling of ultramassive white dwarfs, because it is exacerbated by Debye cooling (solid lines). 

\subsection{Relation to observations}

\begin{figure}
    \centering
    \includegraphics[width=\columnwidth]{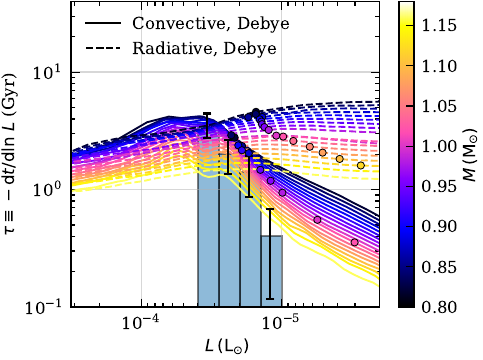}
    \caption{The instantaneous cooling time $\tau$ for our entire grid of CO white dwarfs (colour coded by mass) with hydrogen atmospheres and the standard Debye heat capacity. Solid curves account for convective coupling, whereas dashed curves represent fully radiative models, mimicking the inhibition of convective coupling by a strong magnetic field. Circular markers indicate a total cooling time of $t=10\textrm{ Gyr}$ for each model. The histogram and error bars represent the luminosity distribution of the IR-faint white dwarfs in the \citet{Bergeron2022} sample with masses $M>0.8\,\msun$ (see text).}
    \label{fig:colorplot}
\end{figure}

We emphasize the difference between convectively coupled and radiatively detached envelopes in Fig. \ref{fig:colorplot}, where we plot the cooling curves for our entire range of massive CO white dwarfs with $0.8\,\msun\lesssim M \lesssim 1.16 \,\msun$. We overlay these curves with a histogram (normalized at the peak to our nominal $1.0\,\msun$ curve) of the IR-faint white dwarfs from the \citet{Bergeron2022} sample that exceed $M>0.8 \,\msun$. This model-independent cut corresponds (up to a small margin, see Appendix \ref{sec:appendixmixed}) to the subset of the sample that is in the Debye regime \citep[see fig. 18 of][]{Bergeron2022}, whereas the statistics of lower-mass stars is unaffected by Debye cooling. The histogram error bars assume a Poisson distribution. 

Analysing potential biases in the \citet{Bergeron2022} sample is beyond the scope of this paper. Moreover, $\tau(L)$ is only a proxy for the luminosity function \citep{Fontaine2001,GarciaBerroOswalt2016}, since it ignores the initial mass function, the star formation rate, and the finite age of the Galactic disc in particular, which we approximately account for by marking the $t=10\textrm{ Gyr}$ \citep{Gianninas2015,Cukanovaite2023,Roberts2025,Roberts2026} total cooling time for each curve. Comparing our $\tau(L)$ to the histogram is further complicated by the possible merger origin of many of the ultramassive white dwarfs \citep{Jewett2024}, which effectively resets their cooling clock, introducing a merger delay-time distribution \citep{Maoz2012,Maoz2018}.\footnote{Neon distillation may also stop the cooling clock for several Gyrs, but only for a small fraction of the ultramassive white dwarf population \citep{Cheng2019,Bedard2024}.} Therefore, while the drop in the number of observed ultramassive white dwarfs at low luminosities is consistent with their accelerated Debye cooling (Fig. \ref{fig:colorplot}), as suggested by \citet{Kilic2025,Kilic2026}, other factors such as the maximal cooling time and incompleteness of faint stars might be contributing to this drop as well.  

None the less, Fig. \ref{fig:colorplot} demonstrates that samples similar to \citet{Bergeron2022} may probe the physics of Debye cooling. Specifically, the appreciable difference between the convective and radiative curves even at $t<10\textrm{ Gyr}$ (for ultramassive white dwarfs $M\gtrsim 1.0\,\msun$) may constrain the role of magnetic fields in inhibiting convective coupling. Interestingly, the only ultramassive DA white dwarf in the Debye regime found by \citet{Kilic2026} is magnetic, consistent with the almost flat $\tau(L)$ of strongly magnetized white dwarfs, compared to the steeply shortening cooling time of non-magnetic stars. 

The different slopes of $\tau(L)$ for $10^{-5}\,\lsun\lesssim L\lesssim 10^{-4}\,\lsun$ in Fig. \ref{fig:debye} indicate that samples like \citet{Bergeron2022} could also probe the microphysics of the phononic heat capacity. Robust conclusions, however, would probably require larger samples that extend to lower luminosities and focus on the most massive white dwarfs, which reach such luminosities faster (see Fig. \ref{fig:colorplot}). 

\section{Conclusions}
\label{sec:conclusions}

Recent observations of cold and massive white dwarfs have revealed -- for the first time -- populations that are expected to be in the Debye cooling regime \citep{Bergeron2022}: an accelerated cooling phase driven by the vanishing solid heat capacity $C_V$ at very low internal temperatures $T$ \citep{MestelRuderman1967,LambVanHorn1975}. Here, we analysed the cooling of such white dwarfs, both analytically and numerically, by calculating their instantaneous cooling time $\tau(L)\equiv -\textrm{d} t/\textrm{d}\ln L$, which is a proxy for the luminosity function, i.e. the expected number density of white dwarfs as a function of $L$ \citep[e.g.][]{DAntonaMazzitelli1989,Fontaine2001,GarciaBerroOswalt2016}.

We found that $\tau(L)$ decreases (i.e. cooling accelerates) at $L\lesssim 10^{-4}\,\lsun$ (for DA atmospheres, and somewhat earlier for helium and mixed atmospheres; see Appendix \ref{sec:appendixmixed}) for ultramassive white dwarfs as a combined result of convective coupling and Debye cooling -- which are almost simultaneous -- with neither one of these effects sufficient to significantly accelerate the cooling on its own. Specifically, Debye cooling amplifies the effect of convective coupling (and vice versa), which is otherwise a relatively minor event, similarly to normal-mass white dwarfs which never reach the Debye regime \citep[e.g.][]{Ginzburg2024}. Conversely, strong magnetic fields, which are common in cold and in massive white dwarfs \citep{BagnuloLandstreet2022,Moss2025}, and may inhibit convective coupling \citep{Tremblay2015,Ginzburg2024}, have a much bigger effect on ultramassive white dwarfs, due to their Debye heat capacity. Such magnetic ultramassive white dwarfs should follow an almost flat $\tau\propto L^{1/7}$ distribution for $L\lesssim 10^{-4}\,\lsun$.

We compared our results to a subset of 39 massive ($M>0.8\,\msun$) white dwarfs from the \citet{Bergeron2022} sample that are likely in the Debye regime. While the drop in their number density with $L$ is consistent with the Debye cooling of non-magnetic (i.e. convectively coupled) white dwarfs, other factors may contribute to this drop, such as the finite age of the Galactic disc or observational biases. None the less, our magnetic and non-magnetic $\tau(L)$ curves diverge appreciably at cooling ages $t<10\textrm{ Gyr}$ for ultramassive ($M\gtrsim 1.0\,\msun$) white dwarfs, such that larger samples focusing on this mass range could constrain the magnetic inhibition of convective coupling. 

In principle, measuring $\tau(L)$ for a sufficiently large sample could also constrain the microphysics of the solid heat capacity $C_V(T)$. We parametrized the uncertainty in $C_V$ (e.g. the phonon spectrum of CO and ONe binary-component plasmas) by generalizing the \citet{Chabrier1992} expression for the free energy of harmonic phonons in the solid lattice. Specifically, we introduced a free parameter $\delta$ that sets the low-temperature limit of the phonon heat capacity $C_V\propto T^{\delta}$, with $\delta=3$ corresponding to the standard Debye model. We found that the acceleration of the white dwarf's cooling -- i.e. the slope of $\tau(L)$ -- is sensitive to $\delta$ if the heat capacity vanishes more gradually ($\delta<3$). 
Models with a steeply vanishing phonon heat capacity ($\delta>3$), on the other hand, converge onto the same cooling track as the standard Debye model, because their total heat capacity becomes dominated by the degenerate electrons ($C_V\propto T$). 

In summary, observations are beginning to probe the Debye cooling phase, which has so far been explored only theoretically. By comparing the observed luminosity function to the theoretical cooling time through each luminosity bin $\tau(L)$ in this regime, we may constrain the role of magnetic fields in inhibiting convective coupling, as well as the microphysics of solid ionic mixtures at low temperatures. In this context, \citet{CamisassaEx} recently studied the exsolution process in such mixtures, finding that only 7 stars in the \citet{Bergeron2022} sample are in the Debye regime, due to changes in white dwarf cooling caused by this process. More theoretical work and larger observational samples may shed light on the interplay between exsolution and Debye cooling.

\section*{Acknowledgements}

We thank Michal Isaac and Rom Yaakovyan for useful discussions. This research was partially supported by the United States-Israel Binational Science Foundation (BSF; grant no. 2022175), the German-Israeli Foundation for Scientific Research and Development (GIF; grant no. I-1567-303.5-2024), and the Israel Science Foundation (ISF; grant nos 1600/24 and 1965/24).

\section*{Data Availability}

The data underlying this article will be shared on reasonable request to the corresponding author.



\bibliographystyle{mnras}
\bibliography{debye} 




\appendix

\section{Helium and mixed atmospheres}\label{sec:appendixmixed}

Before entering the Debye regime ($T_{\rm eff}\gtrsim 5000\textrm{ K}$) -- when the H$\alpha$ line is visible and spectroscopy is reliable -- the vast majority of ultramassive white dwarfs have hydrogen-dominated (DA) atmospheres \citep{Kilic2025}. The spectra in the \citet{Bergeron2022} sample, on the other hand, are best explained by mixed helium-dominated outer atmospheres with trace abundances of hydrogen, typically in the $-4.5\lesssim \log(\textrm{H/He})\lesssim -2.5$ range. This apparent discrepancy could be reconciled, in principle, by thin pure hydrogen envelopes that are later mixed by convection as the white dwarfs cool down and their surface convection zones penetrate into the underlying helium layers \citep[see][for a review]{bedard2024review}. In practice, \citet{Bergeron2022} computed the hydrogen envelope mass $M_{\rm H}$ that is required to reproduce the observed abundances in such a scenario and the effective temperature $T_{\rm eff}(M_{\rm H})$ when convective mixing occurs -- finding that more non-DA stars should have been observed at high $T_{\rm eff}$. This remaining discrepancy might indicate that there are large uncertainties in determining $\log(\textrm{H/He})$ from the IR-faint spectrum, or in theoretically computing the depth of the convection zone as a function of $T_{\rm eff}$ for such mixed atmospheres.  

We use the \texttt{replace\_element} procedure in \textsc{mesa} to replace all or part of the hydrogen envelope of a $1.0\,\msun$ CO white dwarf, creating two models: a pure helium atmosphere, and a thin hydrogen outer layer with $M_{\rm H}\approx 3\times 10^{-10}\,\msun$, which is later convectively mixed -- producing a mixed atmosphere with $\log(\textrm{H/He})\approx -4.3$.\footnote{The helium-atmosphere model is cooled using the relevant DB tabulated outer boundary conditions in \textsc{mesa}, whereas for the mixed-atmosphere model we use the \texttt{T\_tau (Eddington, iterated)} option instead.}
While our mixed atmosphere is produced similarly to \citet{Bergeron2022}, the relation between the initial $M_{\rm H}$ and the final $\log(\textrm{H/He})$ is different, due to differences between our evolutionary codes and outer boundary conditions, and the size of the convection zone as a function of $T_{\rm eff}$ in particular \citep[see][]{Rolland2018}.

\begin{figure}
    \centering
    \includegraphics[width=\columnwidth]{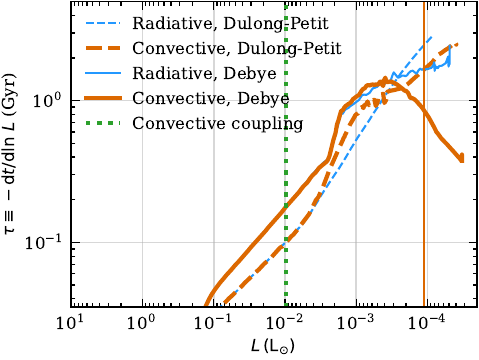}
    \newline
    \includegraphics[width=\columnwidth]{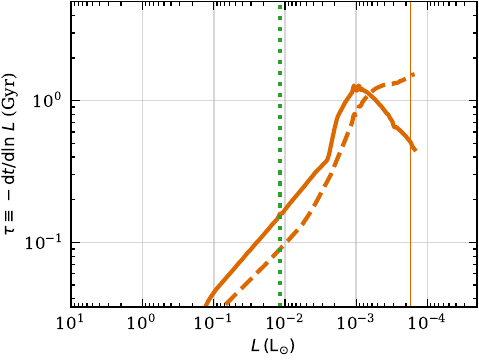}
    \caption{Same as Fig. \ref{fig:convection}, but for CO white dwarfs with a pure helium atmosphere (top panel) and a mixed atmosphere with $\log(\textrm{H/He})\approx -4.3$ (bottom panel, note that convection is necessary to mix the envelope). The vertical dotted line indicates the onset of convective coupling.}
    \label{fig:appendix}
\end{figure}

As seen in Fig. \ref{fig:appendix}, our main conclusions are valid for such helium and mixed atmospheres as well. Specifically, the accelerated cooling at low luminosities is a combined effect of convective coupling and the Debye heat capacity. Moreover, the slope of this acceleration $\tau(L)$ is similar to the DA case, justifying our comparison of the \citet{Bergeron2022} sample to DA atmospheres in Fig. \ref{fig:colorplot}.

The main difference compared to DA atmospheres is the higher luminosity at which this acceleration begins, i.e. the peak of $\tau(L)$. Similarly to \citet{Althaus2007}, our pure helium and mixed atmospheres reach the adopted Debye threshold $T_{\rm D}/T=20$ at the centre at somewhat higher $L$ compared to pure hydrogen atmospheres (solid vertical lines in Figs \ref{fig:convection} and \ref{fig:appendix}). More crucially, while this arbitrary threshold \citep[cf.][]{Althaus2010} coincides with the $\tau(L)$ peak for DA atmospheres, it significantly underestimates $L$ at the peak -- the effective start of the accelerated Debye cooling -- for helium and mixed atmospheres.
The reason for this is the interplay between Debye cooling and convective coupling. Helium and mixed atmospheres develop convection at higher $T_{\rm eff}$ \citep{Rolland2018,Bedard2024}, such that convective coupling occurs at much higher luminosities (vertical dotted lines in Fig. \ref{fig:appendix} compared to Fig. \ref{fig:stagescooling}). As explained analytically in Section \ref{sec:cooling}, this early convective coupling amplifies the small differences between the Debye and Dulong--Petit heat capacities that already exist even at $T_{\rm D}/T<20$ (Fig. \ref{fig:cv}), when (in addition) only the highest-density central regions of the star have cooled much below $T_{\rm D}\propto\rho^{1/2}$ \citep[see also][]{DAntonaMazzitelli1989}. The earlier effective onset of Debye cooling justifies our inclusion of IR-faint white dwarfs with $0.8\,\msun<M<0.9\,\msun$ in Fig. \ref{fig:colorplot} -- IR-faint white dwarfs with lower masses than  predicted by the DA models in fig. 18 of \citet{Bergeron2022} are undergoing Debye cooling.


\bsp	
\label{lastpage}
\end{document}